\documentclass[aps,showkeys,twocolumn,preprintnumbers,amsmath,amssymb,superscriptaddress,floatfix,nofootinbib]{revtex4}
\usepackage{graphicx,color,dcolumn,booktabs,bm}
\usepackage{amsmath}
\usepackage{graphicx}
\usepackage{longtable,lscape}
\usepackage{txfonts}
\usepackage{overpic}
\usepackage{epsfig}
\usepackage{amssymb}
\usepackage{bbding}
\usepackage{rotating}
\usepackage{epstopdf}
\usepackage{appendix}
\usepackage{indentfirst}
\usepackage{feynmf}   
\usepackage{slashed}  
\usepackage{cases}
\usepackage{color}
\usepackage{multirow}
\usepackage{graphicx,color,dcolumn,booktabs,bm}
\usepackage{cases}
\usepackage{array}

\graphicspath{{Figures/}} %

\usepackage[colorlinks, citecolor=blue,anchorcolor=red,menucolor=red, linkcolor=red,filecolor=red,runcolor=red,urlcolor=blue,frenchlinks=red]{hyperref}

\begin{document}
\title{Understanding the low-lying $\Omega_c$ structures from a coupled-channel perspective}
\author{Ying Zhang}

\affiliation{Department of Physics, Hunan Normal University, Changsha 410081, China}
\affiliation{Key Laboratory of Low-Dimensional Quantum Structures and Quantum Control of Ministry of Education, Changsha 410081, China}
\affiliation{Key Laboratory for Matter Microstructure and Function of Hunan Province, and Hunan Research Center of the Basic Discipline for Quantum Effects and Quantum Technologies, Hunan Normal University, Changsha 410081, China}

\author{Qing-Fu Song}

\affiliation{School of Physics, Central South University, Changsha 410083, China}

\author{Qi-Fang L\"{u}}\email{lvqifang@hunnu.edu.cn}

\affiliation{Department of Physics, Hunan Normal University, Changsha 410081, China}
\affiliation{Key Laboratory of Low-Dimensional Quantum Structures and Quantum Control of Ministry of Education, Changsha 410081, China}
\affiliation{Key Laboratory for Matter Microstructure and Function of Hunan Province, and Hunan Research Center of the Basic Discipline for Quantum Effects and Quantum Technologies, Hunan Normal University, Changsha 410081, China}

\author{Hideko Nagahiro} \email{nagahiro@rcnp.osaka-u.ac.jp} 

\affiliation{Department of Physics, Nara Women’s University, Nara 630-8506, Japan}

\affiliation{Research Center for Nuclear Physics (RCNP), Ibaraki, Osaka 567-0047, Japan}

\author{Atsushi Hosaka}\email{hosaka@rcnp.osaka-u.ac.jp}
\affiliation{Research Center for Nuclear Physics (RCNP), Ibaraki, Osaka 567-0047, Japan}
\affiliation{Advanced Science Research Center, Japan Atomic Energy Agency, Tokai, Ibaraki 319-1195, Japan}
\affiliation{RIKEN Nishina Center 
for Accelerator-Based Science, 2-1, Hirosawa, Wako, Saitama 351-0198, Japan}

\begin{abstract}
We perform a systematic analysis of the low-lying $\Omega_c$ structures in a coupled-channel approach. The couplings between meson-baryon channels, $\Xi_c \bar K$, $\Xi_c^{\prime} \bar K$, $\Xi_c^{*} \bar K$, $\Omega_c\eta$ and $\Omega_c^{*}\eta$, and three-quark bare states $\Omega_c(1P_\lambda)$ are considered. We predict a bound state $\Omega_c(2954)$ below the $\Xi_c \bar K$ threshold with $(J^P,j)=(1/2^-,0)$, which can be studied in the final states of $\Omega_c^{(*)} \pi$ and $\Omega_c^{(*)} \gamma$. Also, the resonances $\Omega_c(3000)$ and $\Omega_c(3050)$ can be classified as the lower $(J^P,j)=(1/2^-,1)$ and $(J^P,j)=(3/2^-,1)$ states, respectively. Our present assignments based on this coupled-channel perspective are significantly different from those of traditional three-quark picture and of molecular scenario. The future BelleII and LHCb experiments can search for the bound state $\Omega_c(2954)$ and measure the spin-parities of particles $\Omega_c(3000)$ and $\Omega_c(3050)$ to test our predictions.
\end{abstract}


\maketitle
\section{introduction}

In the past decade, significant achievement has been made in the heavy baryon sector by experimentalists and theorists, which deepens our understanding of the mass spectra and internal structures of heavy-light systems~\cite{Chen:2016spr,Cheng:2021qpd,Chen:2022asf,Nieves:2024dcz,Crede:2024hur}. Although the traditional three-quark picture can well explain many heavy hadrons in experiments, some newly observed resonances can be hardly accommodated into the conventional baryons and more exotic scenarios are necessary. The exotic interpretations mainly include the weakly bound hadronic molecules, compact multiquark states, hybrids, and so on. Meanwhile, the superposition of conventional hadrons and pure exotic states is also possible, known as the unquenched picture. Nowadays, exploring the coupled-channel mechanism in the unquenched picture and clarifying the nature of new hadronic states is an intriguing and challenging topic in hadron physics~\cite{Tornqvist:1984fy,Silvestre-Brac:1991qqx,Hwang:2004cd,Pennington:2007xr,Barnes:2007xu,Li:2009ad,Zhou:2011sp,Liu:2011yp,Nagahiro:2014mba,Nagahiro:2011jn,Nagahiro:2013hba,Yang:2021tvc,Zhang:2022pxc,Ortega:2009hj,Liu:2016wxq,Ferretti:2014xqa,Lu:2017hma,Luo:2019qkm,Luo:2021dvj,Xie:2021dwe,Ni:2021pce,Yamaguchi:2019vea,Lu:2022puv}. Actually, the heavy baryons, such as the $\Omega_c$ family, can provide an ideal platform to investigate this unquenched dynamical mechanism and help us to establish and complete the hadron spectroscopy of heavy-light systems. 

In 2017, the LHCb Collaboration observed five narrow $\Omega_{c}^0$ excited states in the $\Xi_{c}^+K^-$ invariant mass spectrum, which are labeled as $\Omega_{c}(3000)^0$, $\Omega_{c}(3050)^0$, $\Omega_{c}(3066)^0$, $\Omega_{c}(3090)^0$, and $\Omega_{c}(3119)^0$~\cite{LHCb:2017uwr}. Subsequently, the Belle Collaboration confirmed the existence of four structures $\Omega_{c}(3000)^0$, $\Omega_{c}(3050)^0$, $\Omega_{c}(3066)^0$, and $\Omega_{c}(3090)^0$~\cite{Belle:2017ext}. Also, the experimental data of LHCb and Belle Collaborations suggested the presence of a broad structure around 3188 MeV. In 2021, the LHCb Collaboration performed the analysis of helicity angular distributions for resonances $\Omega_{c}(3000)^0$, $\Omega_{c}(3050)^0$, $\Omega_{c}(3065)^0$, and $\Omega_{c}(3090)^0$ in the $\Omega_b^- \to \Xi_c^+ K^- \pi^-$ decay~\cite{LHCb:2021ptx}. They found that the $J=1/2$ spin hypothesis for $\Omega_{c}(3050)^0$ and $\Omega_{c}(3065)^0$ states was excluded with significance of $2.2\sigma$ and $3.6 \sigma$, respectively. Also, the combined hypothesis on the spin of the four peaks, $\Omega_{c}(3000)^0$, $\Omega_{c}(3050)^0$, $\Omega_{c}(3065)^0$, and $\Omega_{c}(3090)^0$, in the order $J=1/2$, $1/2$, $3/2$, and $3/2$ was rejected with a $p-$value corresponding to 3.5 standard deviations. Moreover, an enhancement near the $\Xi_c \bar K$ threshold was seen in both the inclusive $\Xi_c^+ K^-$ spectrum and exclusive $\Omega_b^- \to \Xi_c^+ K^- \pi^-$ decay. In 2023, the LHCb Collaboration further studied the $\Xi_c^+ K^-$ invariant mass spectrum
using proton-proton collision data and observed two new resonances $\Omega_c(3185)$ and $\Omega_c(3327)$ that locate near the $\Xi D$ and $\Xi D^*$ thresholds, respectively~\cite{LHCb:2023sxp}. The measured masses and widths for these structures are collected in Table~\ref{exp} for reference.
\begin{table}[!htbp]
	\begin{center}
		\caption{\label{exp} The experimental information including masses and widths for the observed $\Omega_c$ structures~\cite{LHCb:2023sxp}. It should be noted that the name $\Omega_{c}(3066)^0$ was first used in Refs.~\cite{LHCb:2017uwr,Belle:2017ext}, while the name $\Omega_{c}(3065)^0$ was adopted in the subsequent experiments~\cite{LHCb:2021ptx,LHCb:2023sxp}. In the present work, we employ consistently the notation $\Omega_{c}(3065)^0$ to avoid confusion.}
		\renewcommand{\arraystretch}{1.2}
		\small
		\begin{tabular*}{8.5cm}{@{\extracolsep{\fill}}p{2.0cm}<{\centering}p{3.5cm}<{\centering}p{2.5cm}<{\centering}}
			\hline\hline
 States & Mass($\mathrm{MeV}$)  &  Width$(\mathrm{MeV}$) \\
\hline 
 $\Omega_{c}(3000)^{0}$  &  $3000.44 \pm 0.07^{+0.07}_{-0.13}\pm 0.23$ &  $3.83 \pm 0.23^{+1.59}_{-0.29}$  \\
 $\Omega_{c}(3050)^{0}$  &  $3050.18 \pm 0.04^{+0.06}_{-0.07}\pm 0.23$  &  $0.67 \pm 0.17^{+0.64}_{-0.72}$  \\
 $\Omega_{c}(3065)^{0}$  &  $3065.63 \pm 0.06^{+0.06}_{-0.06}\pm 0.23$  &  $3.79 \pm 0.20^{+0.38}_{-0.47}$    \\
 $\Omega_{c}(3090)^{0}$  &  $3090.16 \pm 0.11^{+0.06}_{-0.10}\pm 0.23$  &  $8.48 \pm 0.44^{+0.61}_{-1.62}$    \\
 $\Omega_{c}(3119)^{0}$  &  $3118.98 \pm 0.12^{+0.09}_{-0.23}\pm 0.23$  &  $0.60 \pm 0.63^{+0.90}_{-1.05}$  \\
 $\Omega_{c}(3185)^{0}$  &  $3185.1 \pm 1.7^{+7.4}_{-0.9}\pm 0.2$  &  $50 \pm 7^{+10}_{-20}$   \\
 $\Omega_{c}(3327)^{0}$  &  $3327.1 \pm 1.2^{+0.1}_{-1.3}\pm 0.2$  &  $20 \pm 5^{+13}_{-1}$   \\
\hline \hline
		\end{tabular*}
	\end{center}
\end{table}

These experimental discoveries have stimulated wide interests among theorists, and plenty of works have been done in the past few years. Within the constituent quark models, lattice QCD calculations, QCD sum rule, effective field theory, and Regge approach, researchers attempted to classify these resonances as conventional charmed-strange baryons~\cite{Chen:2017sci,Karliner:2017kfm,Wang:2017hej,Wang:2017vnc,Padmanath:2017lng,Cheng:2017ove,Wang:2017zjw,Zhao:2017fov,Chen:2017gnu,Agaev:2017lip,Wang:2017kfr,Chen:2017fcs,Santopinto:2018ljf,Gandhi:2019bju,Faustov:2020gun,Bahtiyar:2020uuj,Jia:2020vek,Chen:2021eyk,Polyakov:2022eub,Garcia-Tecocoatzi:2022zrf,Yu:2022ymb,Yu:2023bxn,Luo:2023sra,Oudichhya:2023awb,Ortiz-Pacheco:2023kjn,Pan:2023hwt,Peng:2024pyl,Li:2024zze,Weng:2024roa,Ortiz-Pacheco:2024qcf,Zhong:2025oti,Garcilazo:2007eh,Ebert:2011kk,Wang:2023wii}. Also, the exotic interpretations including meson-baryon molecules and compact pentaquark states were performed in the literature~\cite{Yang:2017rpg,Huang:2017dwn,Kim:2017jpx,Liu:2017frj,An:2017lwg,Kim:2017khv,Montana:2017kjw,Debastiani:2017ewu,Wang:2017smo,Chen:2017xat,Nieves:2017jjx,Huang:2018wgr,Debastiani:2018adr,Wang:2018alb,Xu:2019kkt,Zhu:2022fyb,Yan:2023tvl,Ozdem:2023okg,Wang:2023eng,Ikeno:2023uzz,Xin:2023gkf}. Furthermore, considering the similarity of charmed-strange baryons and mesons, the authors in Ref.~\cite{Luo:2021dvj} investigated the coupled-channel effects by using the $^3P_0$ model in an unquenched picture. 

Various previous studies have offered different interpretations, which can only explain the current experimental measurements partly and are far from reaching a consensus. For instance, in the conventional quark model, the five prominent peaks of $\Omega_c(1P_\lambda)$ states are expected to distribute in a rather narrow mass range, typically some tens of MeV~\cite{Roberts:2007ni,Yoshida:2015tia,Chen:2017gnu,Yu:2022ymb,Luo:2023sra,Peng:2024pyl}, which is significantly smaller than the experimental value $M[\Omega_c(3119)]-M[\Omega_c(3000)] = 119~\rm MeV$. Also the naive spin assignment when the five states are identified with one of $\Omega_c(1P_\lambda)$ state has been ruled out by experiments~\cite{LHCb:2021ptx}. Moreover, whether the $\Omega_c(3119)$ resonance belongs to the $\Omega_c(1P)$ or $\Omega_c(2S)$ state is still widely disputed~\cite{Cheng:2021qpd}. Furthermore, the threshold enhancement observed by experiments may imply the existence of a subthreshold structure in the $\Xi_c \bar K$ system, which is usually ignored in the past studies. It can be seen that the understanding of the mass spectrum and internal structures for $\Omega_{c}$ family is still poor, especially for the unquenched mechanism. In theory, since these observed low-lying $\Omega_c$ states lie in the mass region of $\Xi_c \bar K$, $\Xi_c^\prime \bar K$, and $\Xi_c^{*} \bar K$ thresholds, the coupled-channel effects should be essential for describing their properties, which is analogous to the charmed-strange particle $D_{s0}^*(2317)$.  Indeed, this unquenched picture may play a key role in understanding the properties of the near threshold hadrons. Thus, it is time to carefully study the coupled-channel effects between the meson-baryon configurations and three-quark cores for these low-lying $\Omega_c$ structures.

In this work, we perform a systematic analysis of the low-lying $\Omega_{c}$ structures in a coupled-channel approach, where the couplings between the meson-baryon channels and three-quark bare states $\Omega_c(1P_\lambda)$ are investigated. This approach has been employed in the study of the nature of light hadrons $a_1(1260)$, $\sigma$, and $\Omega(2012)$~\cite{Nagahiro:2011jn,Nagahiro:2013hba,Nagahiro:2014mba,Lu:2022puv}, which suggests that both composite and bare components are crucial for describing the properties of these states. Here the composite components correspond to those dynamically generated states via the relevant hadron-hadron interactions. This unified treatment can be extended to the heavy baryons and suitable for investigating the nature of observed low-lying $\Omega_c$ structures. Our results predict a bound state $\Omega_c(2954)$ below the $\Xi_c \bar K$ threshold with $(J^P,j)=(1/2^-,0)$, which can be hunt for in the $\Omega_c^{(*)})\gamma$ and $\Omega_c^{(*)} \pi$ final states by the future Belle II and LHCb experiments. Also, the resonances $\Omega_c(3000)$ and $\Omega_c(3050)$ can be classified as the lower $(J^P,j)=(1/2^-,1)$ and $(J^P,j)=(3/2^-,1)$ states.    

This paper is organized as follows. In Sec.~\ref{formalism}, the relevant effective interactions are derived and the framework of coupled-channel approach is introduced as well. The numerical results and discussions for the present calculations are shown in Sec.~\ref{results}. A brief summary is given in the last section.

\section{Formalism}\label{formalism}
\subsection{The contact interactions for meson-baryon channels}
Observing that the conventional baryons include two light quarks, let us first consider the interaction between a pseudoscalar meson and a light quark which is given by the Weinberg-Tomozawa term in the chiral SU(3) Lagrangian
\begin{eqnarray}
{\mathcal L}_{WT}=-\frac{i}{8f_{p}^2}\bar{q}\gamma^\mu(\phi^\mu\phi-\phi\phi^\mu)q,
   \label{L_WT}
\end{eqnarray}
with
\begin{eqnarray}
	\phi \equiv \vec{\lambda} \cdot \vec{M_p},  \nonumber \\
	\phi^\mu \equiv \vec{\lambda} \cdot \partial^\mu \vec{M_p}.
\end{eqnarray}
Here, the $\vec{M_p}$ is the light pseudoscalar meson field, $\vec{\lambda}$ is the SU(3) Gell-Mann matrix, and $f_p$ corresponds to a decay constant for a pseudoscalar meson. 

\begin{figure}
    \centering
\includegraphics[scale=0.65]{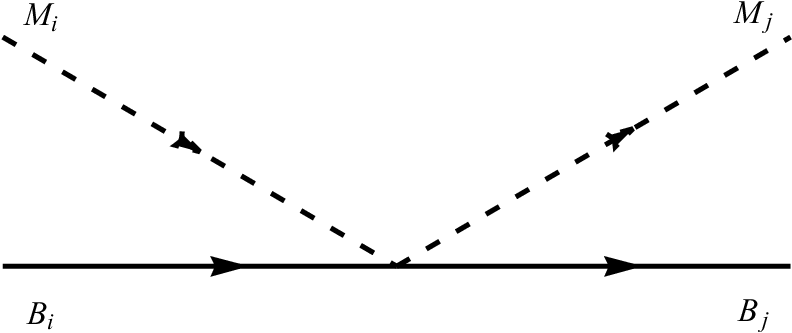}
    \caption{The Feynman diagram for contact interactions for meson-baryon channels.}
    \label{mb1}
\end{figure}

The Feynman diagram for the contact interactions for meson-baryon channels is shown in Figure~\ref{mb1}. Since the present calculations are conducted for near threshold region where the relevant hadrons move with small momenta, the spatial part of interacting fields can be neglected and only the predominating $\partial_{0}$ and $\gamma_{0}$ terms need to be considered. Then, the contact interactions for meson-baryon channels can be written as 
\begin{eqnarray}
   \small
    (V_{\rm com})_{ij}=D_{ij}\frac{2\sqrt{s}-M_{B_i}-M_{B_j}}{4f_{p}^2},
    \label{Vij}
\end{eqnarray}
where $\sqrt{s}$ is the energy for a meson-baryon pair in their center of mass system, $M_{B_i}$ and $M_{B_j}$ represent the masses of interacting baryons. The $D_{ij}$ stand for the flavor factors dependent on the relevant processes and are listed in Table.~\ref{Dij}. Here, we use the $V_{\rm com}$ to represent the matrix for pure meson-baryon interactions, and the $(V_{\rm com})_{ij}$ is the matrix element in the $i-$th row and $j-$th column. It can be seen that these energy dependent interactions at the quark level lead to the same interactions at the hadronic level when the standard quark model wave functions are adopted. 
\begin{table}[!htbp]
	\begin{center}
		\caption{\label{Dij} The relevant flavor factors $D_{ij}$ for the $S-$wave meson-baryon channels. The $J^P$ is the spin-parity, and $j$ stands for the light quark spin of the heavy-light system.}
		\renewcommand{\arraystretch}{1.2}
		\small
		\begin{tabular*}{8.5cm}{@{\extracolsep{\fill}}p{1.1cm}<{\centering}p{1.1cm}<{\centering}p{1.0cm}<{\centering}p{4.5cm}<{\centering}}
			\hline\hline
			$(J^P,j)$ & Channel & \multicolumn{2}{c}{$D_{ij}$}  \\\hline
$(1/2^-,0)$ & $\Xi_{c}\bar{K}$   & \multicolumn{2}{c}{$-1$} \\
\multirow{2}{*}{$(1/2^-,1)$} & \multirow{2}{*}{$\Xi_{c}^\prime\bar{K}$, $\Omega_{c}\eta$}    &  \multirow{2}{*}{$\begin{pmatrix}-1& -4/\sqrt{3} \\-4/\sqrt{3} & 0 \end{pmatrix}$} \\
 &   \\
 \multirow{2}{*}{$(3/2^-,1)$} & \multirow{2}{*}{$\Xi_{c}^{*}\bar{K}$, $\Omega_{c}^*\eta$ }   &  \multirow{2}{*}{$\begin{pmatrix}-1& -4/\sqrt{3} \\-4/\sqrt{3} & 0 \end{pmatrix}$} \\
 &  \\
\hline\hline
		\end{tabular*}
	\end{center}
\end{table}

\subsection{The interactions between meson-baryon channels and three-quark bare states}\label{sec3}

To investigate the coupled-channel effects between the meson-baryon channels and three-quark bare states, we adopt the axial-vector type coupling for the interaction between the pseudoscalar meson and a light quark inside the heavy baryons. The interaction can be expressed as 
\begin{equation}
	\mathcal{L}_{M_{p}qq}=\frac{g_{A}^q}{2f_{p}}\bar{q}\gamma_{\mu}\gamma_{5} \phi^\mu q,
	\label{L_Mqq}
\end{equation}
where $g_{A}^q$ represents the quark-axial-vector coupling strength. This
interaction can be also obtained from the chiral SU(3) Lagrangian and has been extensively employed in the quark model calculations~\cite{Lu:2022puv,Liu:2019wdr,Song:2023cyk,Arifi:2022ntc,Nagahiro:2016nsx,Wang:2017hej,Arifi:2017sac,Shu:2024jdn}. The quark axial-vector coupling can take any value in the non-linear realization of chiral symmetry, while the vector coupling strength in the Weinberg-Tomozawa term (\ref{L_WT}) is uniquely determined. In the present study we employ $g_A^q = 1$ which is the value when the quark belongs to the lowest-dimensional representation $(1/2,0)+(0,1/2)$ of the linear realization of chiral symmetry~\cite{Weinberg:1969hw}.  

To match the nonrelativistic wave functions of heavy baryons, the interacting operator can be expanded in powers of $1/m$, which is sufficient to describe the low momentum phenomena. The  nonrelativistic operator can be written as
\begin{eqnarray}
	\mathcal{H}_{NR}=\frac{g_A^q}{2f_{p}}[\boldsymbol{\sigma} \cdot \boldsymbol{q} + \frac{\omega}{2m} (\boldsymbol{\sigma} \cdot \boldsymbol{q}-2\boldsymbol{\sigma} \cdot \boldsymbol{p_{i}})].
\end{eqnarray}
Here, $m$ is the constituent mass of $u/d/s$ quark by assuming SU(3) flavor symmetry for the light quarks, $\omega$ is the energy of the outgoing pseudoscalar meson, $\boldsymbol{q}$ is the momentum of the outgoing pseudoscalar meson, and $\boldsymbol{p_{i}}$ stands for the momentum of the $i-$th light quark inside the initial baryon. It is noted that only the two light quarks in the heavy baryons couple with the pseudoscalar meson, while the heavy quark acts as a spectator.

\begin{figure}
    \centering
\includegraphics[scale=0.65]{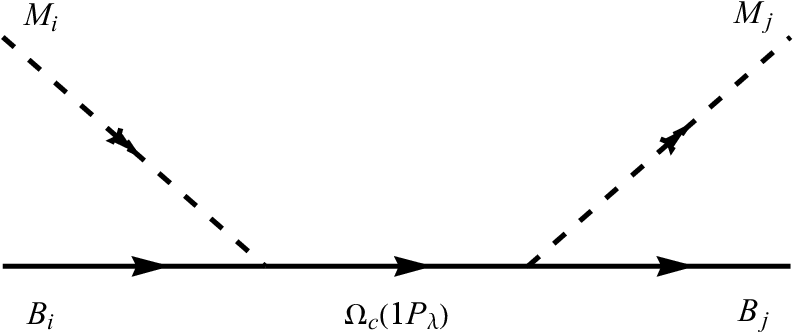}
    \caption{The Feynman diagram for the interactions between meson-baryon channels and three-quark bare states $\Omega_c(1P_\lambda)$.}
    \label{mb2}
\end{figure}

According to the heavy quark spin symmetry, the five $\lambda-$mode $\Omega_c(1P_\lambda)$ states are classified by the total angular momentum of light quark subsystem, which is usually referred to as the light quark spin $j$. The light quark spin $j$ is then combined with the heavy quark spin $1/2$, giving the total angular momentum $J$ of $\Omega_c(1P_\lambda)$ states with $J =1/2$, $3/2$, and $5/2$. Together with the parity $P$, the states are labeled by $(J^P, j)$. Similarly, the meson-baryon channels can be classified as shown in Table~\ref{Dij}. In the coupled-channel scenario, physical states are expressed as superpositions of the two structures, meson-baryon molecular states and the three-quark bare states $\Omega_c(1P_\lambda)$. The coupling scheme between them is shown in Table~\ref{coupling}, and the corresponding Feynman diagram is displayed in Figure~\ref{mb2}.
\begin{table}[!htbp]
	\begin{center}
		\caption{\label{coupling} The coupling scheme between the meson-baryon channels and three-quark bare states. The spin-parity $J^P$ and light quark spin $j$ are adopted to denote a certain system.}
		\renewcommand{\arraystretch}{1.2}
		\small
		\begin{tabular*}{8.5cm}{@{\extracolsep{\fill}}p{2cm}<{\centering}p{3.0cm}<{\centering}p{3.0cm}<{\centering}}
			\hline\hline
			$(J^P,j)$ & Meson-baryon channel &  Three-quark bare state \\\hline
$(1/2^-,0)$ & $\Xi_{c}\bar{K}$   & $\Omega_{c0}(1/2^-)$\\
$(1/2^-,1)$ & $\Xi_{c}^\prime\bar{K}$, $\Omega_{c}\eta$   & $\Omega_{c1}(1/2^-)$  \\
$(3/2^-,1)$ & $\Xi_{c}^{*}\bar{K}$, $\Omega_{c}^*\eta$   &  $\Omega_{c1}(3/2^-)$ \\
$(3/2^-,2)$ & $\cdot \cdot \cdot$   &  $\Omega_{c2}(3/2^-)$ \\
$(5/2^-,2)$ & $\cdot \cdot \cdot$   &  $\Omega_{c2}(5/2^-)$ \\
\hline\hline
		\end{tabular*}
	\end{center}
\end{table}

To derive the effective interactions, we need to calculate the transition amplitudes between the meson-baryon channels and three-quark bare states, such as $\mathcal{M}[\Omega_{c0}(1/2^-) \to \Xi_c \bar K]$. For the spatial part of initial and final three-quark bare states, the harmonic oscillator wave functions are adopted here, which are common for all up, down, and strange quarks under the assumption of SU(3) symmetry. With the above nonrelativistic operator $\mathcal{H}_{NR}$ and wave functions of baryons, one can calculate all the relevant transition amplitudes between the meson-baryon channels and three-quark bare states. Then, the effective interaction $V_{\rm bare}$ for meson-baryon channels can be obtained by eliminating the $\Omega_c(1P_\lambda)$ states
\begin{eqnarray}
    (V_{\rm{bare}})_{ij}&=&\mathcal{M}[(MB)_i\to \Omega_{c}(1P_\lambda)]\frac{2M_0}{s-M_0^2} \nonumber \\  &&\times \mathcal{M}[\Omega_{c}(1P_\lambda) \to (MB)_j],
\end{eqnarray}
where $s$ is the square of energy in the center of mass system, $M_0$ is the bare mass for a intermediate state $\Omega_c(1P_\lambda)$, and the $(MB)_i$ and $(MB)_j$ stand for the corresponding meson-baryon channels. Similar to the cases of contact meson-baryon interactions, we adopt the $V_{\rm bare}$ to represent the matrix for the interactions between meson-baryon channels and three-quark bare states, where the $(V_{\rm{bare}})_{ij}$ is the matrix element in the $i-$th row and $j-$th column. Moreover, in the $(V_{\rm bare})_{ij}$, Gaussian type form factors appear 
(e.g. $\mathcal{M}[(MB)_i\to \Omega_{c}(1P_\lambda)]$), which plays the role of making the relevant loop integrals finite. In
practice, these Gaussian form factors are replaced by the form factor with the sharp cutoff in three-momentum integral with the cutoff parameter $\Lambda$ as shown in Eq.~(\ref{G1}).


\subsection{The amplitudes}\label{sec3}

By summing the interactions for pure meson-baryon channels $V_{\rm com}$ and the interactions between meson-baryon channels and the three-quark bare states $V_{\rm bare}$, the full interaction in the coupled-channel approach is
\begin{eqnarray}
   V_{\rm full} = V_{\rm com} + V_{\rm bare}.
\end{eqnarray}
In the following discussions, we consider two types of scattering amplitudes: the pure composite amplitude $T_{\rm com}$ obtained by using $V_{\rm com}$ and the full amplitude $T_{\rm full}$ obtained by using $V_{\rm full}$, that is,
\begin{eqnarray}
    T_{\rm com}&=&[1-V_{\rm com} G]^{-1}V_{\rm com},\label{Tcom}
\end{eqnarray}
\begin{eqnarray}
    T_{\rm full}&=&[1-V_{\rm full} G]^{-1}V_{\rm full}.\label{Tfull}
\end{eqnarray}
The function $G$ is the diagonal matrix of intermediate meson-baryon loop functions. After regularizing the loop functions with the sharp cutoff method by $\Lambda$, one can obtain 
\begin{eqnarray}
G_{\Xi_c\bar K/\Xi_c^\prime \bar K/\Omega_c\eta}^I(\sqrt{s})&=&\frac{M_{B}}{2\pi^2}\int_{0 }^{\Lambda}dq\frac{q^2}{s-(E_{B}+\omega_{M})^2+i\epsilon} \nonumber\\
	&&\times\frac{E_{B}+\omega_{M}}{E_{B}\omega_{M}},\label{G1}
\end{eqnarray}
\begin{eqnarray}
G_{\Xi_c^{*}\bar K/\Omega_c^*\eta}^I(\sqrt{s})&=&\frac{M_{B}}{2\pi^2}\int_{0 }^{\Lambda}dq\frac{q^2}{s-(E_{B}+\omega_{M})^2+i\epsilon} \nonumber\\
	&&\times\frac{E_{B}+\omega_{M}}{E_{B}\omega_{M}}\bigg(1+\frac{2q^2}{9M_{B}^2}\bigg),\label{G2}
\end{eqnarray}
where the $E_{B}=\sqrt{M_{B}^2+q^2}$ and $\omega_{M}=\sqrt{m_{M}^2+q^2}$ are the energies of intermediate baryons and mesons, respectively. The $q$ is adopted to represent the magnitude of the momentum $\boldsymbol{q}$. The cutoff parameter $\Lambda$ eliminates the ultraviolet divergence, which reflects the finite size of baryons. The term $\big(1+\frac{2q^2}{9M_{B}^2}\big)$ arises from the propagator of spin$-3/2$ baryon~\cite{Lu:2022puv}, which gives minor effects for a small momentum $q$ compared with the baryon mass $M_B$ and is usually omitted in the literature. In the present work, we keep that term. The superscript $I$ stands for the first Riemann sheet.

In addition to the bound states in the first Riemann sheet, we still need to look for poles in the second Riemann sheet. The loop functions in the second Riemann sheet can be defined as 
\begin{eqnarray}
G_{\Xi_c\bar K/\Xi_c^\prime \bar K/\Omega_c\eta}^{II}(\sqrt{s})=G_{\Xi_c\bar K/\Xi_c^\prime \bar K/\Omega_c\eta}^{I}(\sqrt{s})+i\frac{M_Bq}{2\pi\sqrt{s}},
\end{eqnarray}
\begin{eqnarray}
G_{\Xi_c^{*}\bar K/\Omega_c^*\eta}^{II}(\sqrt{s})=G_{\Xi_c^{*}\bar K/\Omega_c^*\eta}^I(\sqrt{s})+i\frac{M_Bq}{2\pi\sqrt{s}}\bigg(1+\frac{2q^2}{9M_{B}^2}\bigg),
\end{eqnarray}
with  
\begin{equation}
	q = \frac{\sqrt{[s-(M_B+m_M)^2][s-(M_B-m_M)^2]}}{2 \sqrt{s}}
\end{equation}
and 
\begin{equation}
	{\rm Im}~q>0.
\end{equation}
By solving the Bethe-Salpeter equations (\ref{Tcom}) and (\ref{Tfull}), we can find the poles of these amplitudes. Moreover, for each pole, one can calculate the coupling strengths $g_i$ between the pole and various meson-baryon channels. The coupling $g_i$ can be extracted from the definition~\cite{Debastiani:2017ewu}
\begin{eqnarray}
    T_{ij}&=&\frac{g_ig_j}{\sqrt{s}-z_R}, \label{gi}
\end{eqnarray}
where $z_R$ is the position of a pole and $T$ can be either $T_{\rm com}$ or $T_{\rm full}$ depending on the information we want to extract, whether we want to know the structure of meson-baryon molecular states, or the states of full coupled channels.

Before doing the numerical calculations, several parameters should be fixed. As we briefly explain all of them are acceptable due to physical reasons and have been tested in various previous studies. The quark-axial-vector coupling constant $g_A^q=1$ is adopted, and the decay constant $f_p=f_K=111~\rm MeV$ is suitable for present multistrangeness baryons. The strength $g_A^q$ would be a bit overestimated~\cite{Nagahiro:2016nsx} which can be partly adjusted by tuning the cutoff parameter $\Lambda$. For the interactions $V_{\rm bare}$, we adopt the average light quark mass $m=450~\rm MeV$, charm quark mass $m_c=1500 ~\rm MeV$, and harmonic oscillator parameter $\alpha_\rho=331~\rm MeV$ in the light subsystem following the previous works~\cite{Arifi:2017sac,Arifi:2022ntc,Lu:2022puv}. Also, the $\alpha_\lambda$ equals to $387~\rm MeV$, which is determined by the relation~\cite{Zhong:2007gp,Liang:2019aag,Liang:2020hbo}
\begin{eqnarray}
    \alpha_\lambda=\alpha_\rho\bigg(\frac{3m_c}{m_c+2m}\bigg)^{1/4}.
\end{eqnarray}
Having fixed these parameters, we will discuss parameter dependence of the amplitudes and poles on the cutoff parameter $\Lambda$ and bare mass $M_0$ in the Sec.~\ref{results}.

\section{Results and discussions}\label{results}  

\subsection{The $j=0$ system}
\begin{figure*}
	\centering
\includegraphics[scale=0.65]{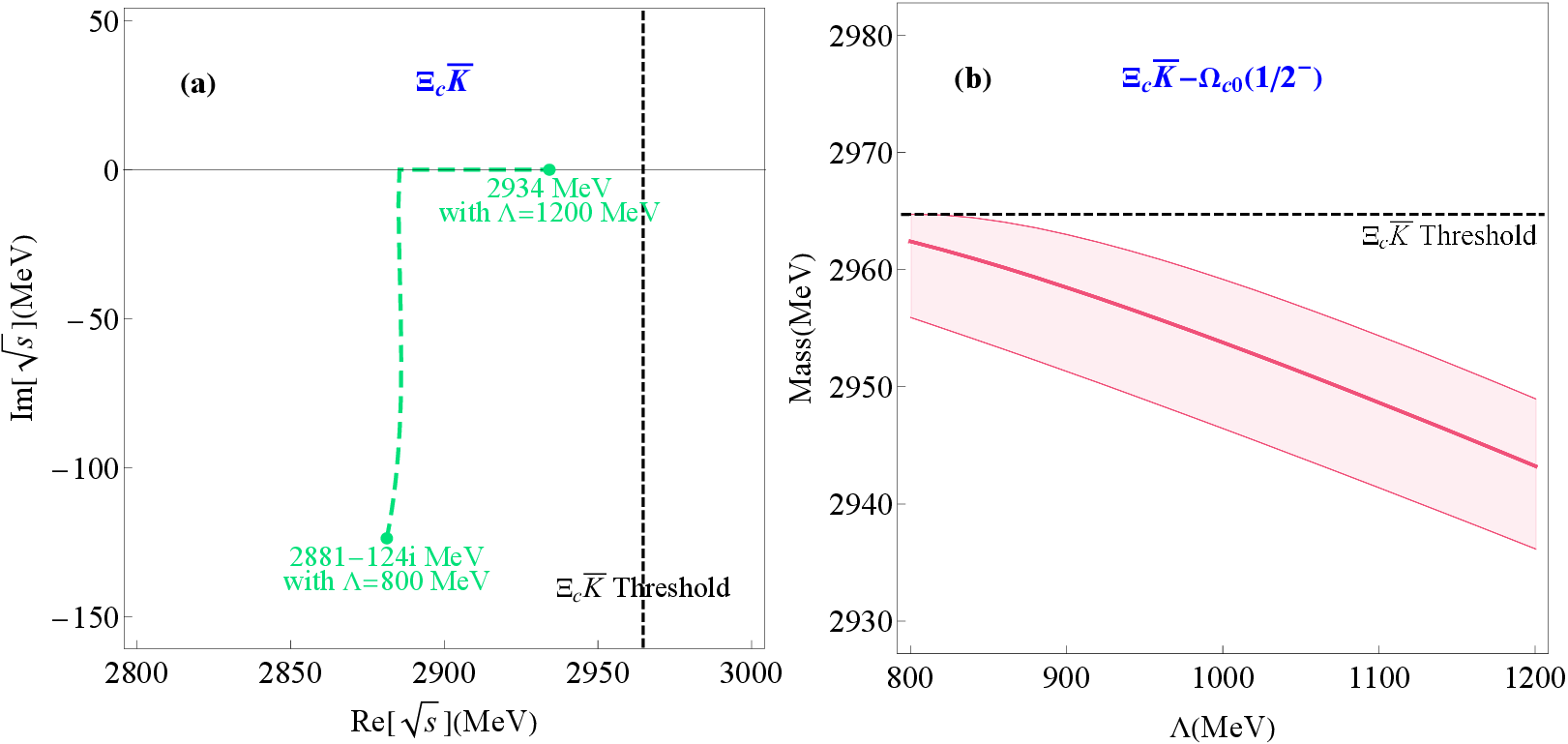}
	\caption{(a) Trajectory of the pole of a pure meson-baryon molecular state with $(J^P,j) = (1/2^-,0)$ in the second Riemann sheet as a function of cutoff parameter $\Lambda$.  (b) The mass of the lower superposition state with $(J^P,j) = (1/2^-,0)$ in full coupled channels as a function of cutoff parameter $\Lambda$. The lower, middle and upper lines are the cases with $M[\Omega_{c0}(1/2^-)]=3000$, $3020$, and $3040~\rm MeV$, respectively.}
	\label{mass1}
\end{figure*}

We first consider the pure meson-baryon channel $\Xi_c \bar K$, which is the only meson-baryon channel that has $j = 0$ as shown in Table~\ref{Dij}. By analyzing the amplitude $T_{\rm com}$, we find a pole in the second Riemann sheet below the $\Xi_c \bar K$ threshold. The pole varies from $2881-124i~\rm{MeV}$ to $2934~\rm{MeV}$ with cutoff parameter $\Lambda$ in the range of $800 \text{--} 1200 ~\rm MeV$, which is shown in Figure~\ref{mass1}. Specifically, with $\Lambda = 1000~\rm MeV$, the virtual state is located at $2886 - 71 i~\rm MeV$.  

Now let us consider the case with coupling the meson-baryon channel $\Xi_c \bar K$ and bare $\Omega_{c0}(1/2^-)$ states with the cutoff parameter $\Lambda$ = 1000 MeV and bare mass $M[\Omega_{c0}(1/2^-)] = 3020  ~\rm MeV$. The two poles of virtual state at $2886 - 71 i~\rm MeV$ and the three-quark bare state at $3020$ MeV change completely. We have found one bound state at 2954 MeV, denoted as $\Omega_c(2954)$, which is only slightly below the $\Xi_c \bar K$ threshold at 2965 MeV. This bound state can cause an enhancement near the $\Xi_c \bar K$ threshold, which corresponds to the threshold structure observed by the LHCb
Collaboration~\cite{LHCb:2017uwr,LHCb:2021ptx}. We also found a pole at $3206-118i~\rm MeV$, causing a broad bump or background. By now, there is no relevant information in experiment, and it may be observed in the $\Xi_c \bar K$ final state in the future. The mass of the lower pole with different cutoff parameter $\Lambda$ and bare mass $M[\Omega_{c0}(1/2^-)]$ are also displayed in Figure~\ref{mass1} for reference.

For the bound state $\Omega_c(2954)$, we can also determine its coupling strength to $\Xi_c \bar K$ channel from the amplitude $T_{\rm full}$.  With Eq.~(\ref{gi}), the predicted coupling strength between the bound state $\Omega_c(2954)$ and $\Xi_c \bar K$ channel is $g_{\Omega_c(2954)\Xi_c \bar K}^2=1.912$. Since it lies below all the Okubo-Zweig-Iizuka-allowed (OZI-allowed) strong decay channels, the total decay width is expected to be rather narrow. Theoretically, the isospin broken decay modes $\Omega_c^{(*)} \pi$ and radiative transition $\Omega_c^{(*)} \gamma$ should dominate. This situation also occurs in the charmed-strange mesons; for instance, the particle $D_{s0}^*(2137)$ has significantly lower mass and narrow decay width. Moreover, our predicted mass for this bound state is consistent with the calculations of low-lying $\Omega_c$ states in the unquenched quark model~\cite{Luo:2021dvj}. We highly suggest that the future BelleII and LHCb Collaborations search for this state in the $\Omega_c^{(*)} \pi$ and $\Omega_c^{(*)} \gamma$ final states.  

\subsection{The $j=1$ systems}

\begin{figure*}
	\centering
\includegraphics[scale=0.65]{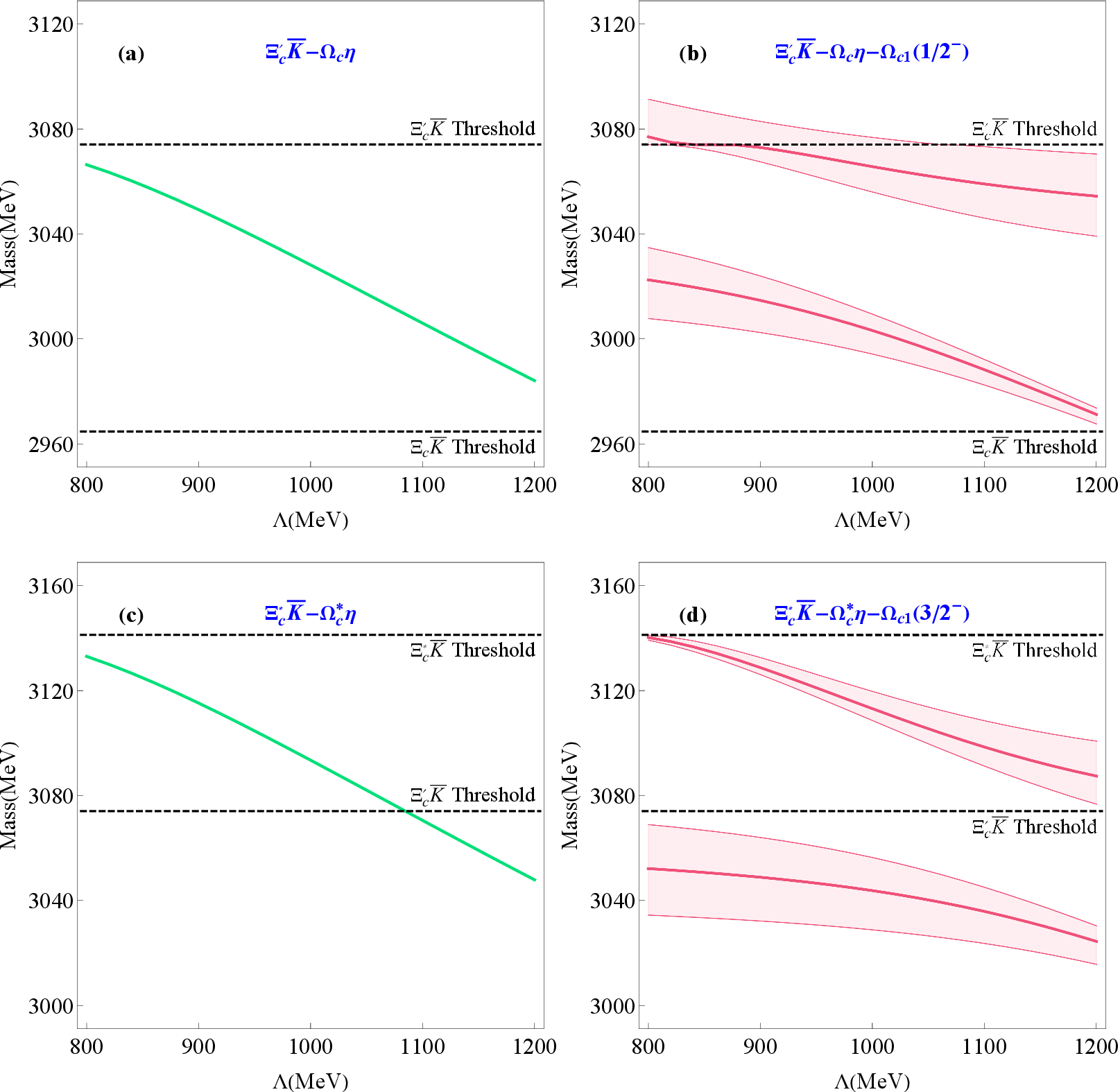}
	\caption{(a) The mass of a pure meson-baryon molecular state with $(J^P,j) = (1/2^-,1)$ as a function of cutoff parameter $\Lambda$.  (b) The masses of the two superposition states with $(J^P,j) = (1/2^-,1)$ in full coupled channels as a function of cutoff parameter $\Lambda$. The lower, middle and upper lines are the cases with $M[\Omega_{c1}(1/2^-)]=3020$, $3040$, and $3060~\rm MeV$, respectively. (c) The mass of a pure meson-baryon molecular state with $(J^P,j) = (3/2^-,1)$ as a function of cutoff parameter $\Lambda$.  (d) The masses of the two superposition states with $(J^P,j) = (3/2^-,1)$ in full coupled channels as a function of cutoff parameter $\Lambda$. The lower, middle and upper lines are the cases with $M[\Omega_{c1}(3/2^-)]=3040$, $3060$, and $3080~\rm MeV$, respectively.}
	\label{mass2}
\end{figure*}

The pure $j=1$ meson-baryon channels include two cases, the $J^P=1/2^-$ $\Xi_c^\prime \bar K-\Omega_c \eta$ and $J^P=3/2^-$ $\Xi_c^{*} \bar K-\Omega_c^* \eta$ configurations. With $\Lambda$ in the range of $800 \text{--} 1200 ~\rm MeV$, we can find a $J^P=1/2^-$ bound state below the $\Xi_c^\prime \bar K$ threshold and a $J^P=3/2^-$ bound state below the  $\Xi_c^{*} \bar K$ threshold. The dependence of cutoff parameter $\Lambda$ is presented in Figure~\ref{mass2}. When the $\Lambda$ varies from 800 to 1200 MeV, the pole of $J^P=1/2^-$ bound state locates in the mass region of $3066 \text{--} 2984~\rm{MeV}$ corresponding to the binding energy of $8 \text{--} 90~\rm{MeV}$. Also, the $J^P=3/2^-$ bound state has a binding energy of $8 \text{--} 93~\rm{MeV}$. Obviously, the binding energies of these two states are quite similar. The reason is that the $\Xi_c^\prime \bar K-\Omega_c \eta$ and $\Xi_c^{*} \bar K-\Omega_c^* \eta$ configurations have similar interactions from the Weinberg-Tomozawa term in the chiral SU(3) Lagrangian and becomes the same in the heavy quark limit.

Then, we add the three-quark bare states $\Omega_{c1}(1/2^-)$ and $\Omega_{c1}(3/2^-)$ into the $(J^P,j)=(1/2^-,1)$ and $(J^P,j)=(3/2^-,1)$ systems, respectively, to investigate the coupled-channel effects between the molecular and three-quark bare states. With the full amplitudes $T_{\rm full}$, we show the pole positions with different cutoff parameter $\Lambda$ and bare masses of $\Omega_{c1}(1/2^-)$ and $\Omega_{c1}(3/2^-)$ in Figure~\ref{mass2}. It can be seen that the two-pole structures emerge in both channels, where the lower poles correspond to bound states, and the higher poles manifest as bound states in the first Riemann sheet or resonances in the second Riemann sheet. Also, when the $\Lambda$ increases from 800 to 1200 MeV, the energies decrease with more binding energy as the attractions increase.

Considering the mass splittings of five $\Omega_c(1P_\lambda)$ states, one expects that the $j=1$ $\Omega_{c1}(1/2^-)$ and $\Omega_{c1}(3/2^-)$ states have larger bare masses than that of $j=0$ $\Omega_{c0}(1/2^-)$ state, and also the bare mass of $\Omega_{c1}(3/2^-)$ state is higher than that of the $\Omega_{c1}(1/2^-)$ state. Thus, we adopt $M[\Omega_{c1}(1/2^-)]=3040~\rm{MeV}$ and $M[\Omega_{c1}(3/2^-)]=3060~\rm{MeV}$. Again let us fix $\Lambda = 1000~\rm MeV$, and look at the coupled-channel effects in more detail. The two systems are the $\Xi_c^\prime \bar K-\Omega_c \eta-\Omega_{c1}(1/2^-)$ and $\Xi_c^{*} \bar K-\Omega_c^* \eta-\Omega_{c1}(3/2^-)$. The two lower poles at 3003 MeV with $J^P=1/2^-$ and at 3044 MeV with $J^P=3/2^-$, are consistent with the masses of observed particles $\Omega_c(3000)$ and $\Omega_c(3050)$, respectively. Although the $\Xi_c \bar K$ channel is OZI-allowed for these poles, the decays into this channel violates the heavy quark spin symmetry. Thus, these poles are expected to be narrow, which is also consistent with the experimental measurements.

\subsection{The whole picture}

As shown in Table~\ref{coupling}, one can find that the $j = 2$ systems have
no coupling to the meson-baryon channels of an $S-$wave. 
Then, in our present model, the $j = 2$ physical states are the three-quark bare states $\Omega_{c2}(3/2^-)$ and $\Omega_{c2}(5/2^-)$. Also, the $j=2$ three-quark states are expected to be a bit higher than the $j=1$ $\Omega_{c1}(1/2^-)$ and $\Omega_{c1}(3/2^-)$ states. The decay widths of these two $j=2$ states are small because of the violation of heavy quark spin symmetry or the necessity of $D-$wave in decay channels. These properties are actually consistent with the observed particles $\Omega_c(3065)$ and $\Omega_c(3090)$. Given the classification of traditional quark model, it is natural to assign the particles $\Omega_c(3065)$ and $\Omega_c(3090)$ as the conventional three-quark states $\Omega_{c2}(3/2^-)$ and $\Omega_{c2}(5/2^-)$, respectively.  

\begin{figure*}
	\centering
\includegraphics[scale=0.80]{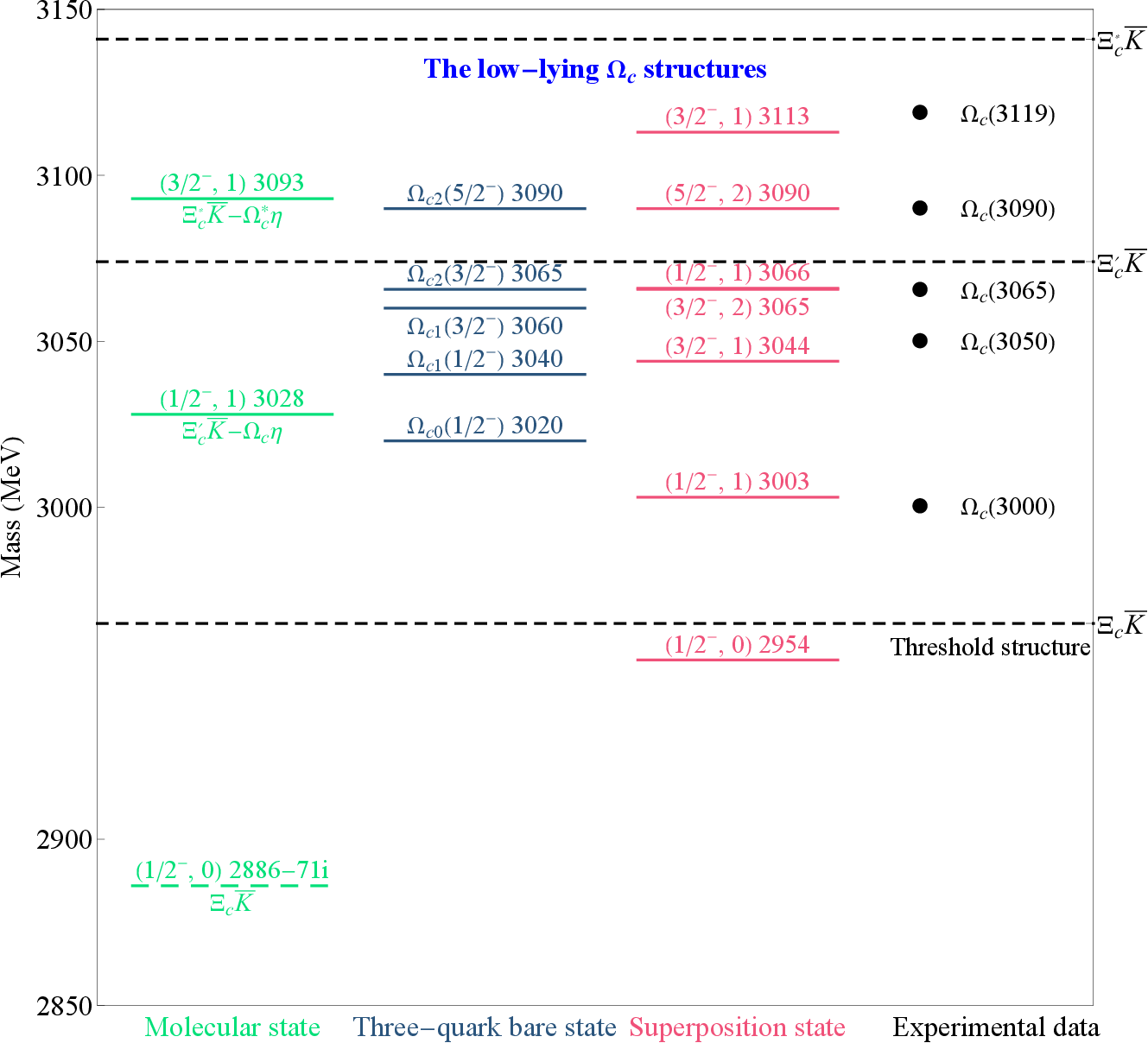}
	\caption{The whole picture for the low-lying $\Omega_c$ structures in the coupled-channel perspective.}
	\label{picture}
\end{figure*}

Now, we get a reasonable picture for the low-lying $\Omega_c$ structures under the coupled-channel approach. First, there are three pure meson-baryon systems, $(J^P,j)=(1/2^-,0)$ $\Xi_c \bar K$, $(J^P,j)=(1/2^-,1)$ $\Xi_c^\prime \bar K-\Omega_c \eta$, and $(J^P,j)=(3/2^-,1)$ $\Xi_c^{*} \bar K-\Omega_c^* \eta$, which generate a virtual state and two bound states. Also, one has five three-quark bare states, $\Omega_{c0}(1/2^-)$, $\Omega_{c1}(1/2^-)$, $\Omega_{c1}(3/2^-)$, $\Omega_{c2}(3/2^-)$, and $\Omega_{c2}(5/2^-)$. Theoretically, the molecular states and three-quark bare states coexist in the low-lying mass region of $\Omega_c$ family. After introducing the couplings between them, we finally obtain seven states in this low-lying energy region except for a $(J^P,j)=(1/2^-,0)$ higher and broad pole. These results together with the experimental data are illustrated in Figure~\ref{picture}. In this scheme, the enhancement near $\Xi_c\bar K$ threshold arising from the undiscovered particle $\Omega_c(2954)$, while the higher $j=0$ pole is too broad to be observed so far. The particles $\Omega_c(3000)$ and $\Omega_c(3050)$ are the lower poles with $(J^P,j)=(1/2^-,1)$ and $(J^P,j)=(3/2^-,1)$, respectively. Also, the two higher poles at 3066 MeV and 3113 MeV may correspond to an unknown particle and the observed state $\Omega_c(3119)$, respectively. Finally, the $\Omega_c(3065)$ and $\Omega_c(3090)$ belong to the traditional $j=2$ $\Omega_c(1P_\lambda)$ states. Furthermore, the relevant couplings between the full coupled-channel states and meson-baryon channels are calculated according to Eq.~(\ref{gi}) and listed in Table~\ref{strength} for reference.

\begin{table}[!htbp]
	\begin{center}
		\caption{\label{strength} The relevant couplings $g_i \times g_j$ between the full coupled-channel states and meson-baryon channels. It should be mentioned that the two $j=2$ states have no coupling to the meson-baryon channels of an $S-$wave.}
		\renewcommand{\arraystretch}{1.2}
		\small
		\centering
		\begin{tabular*}{8.5cm}{@{\extracolsep{\fill}}p{1.7cm}<{\centering}p{1.7cm}<{\centering}p{1.7cm}<{\centering}p{2.4cm}<{\centering}p{5.0cm}<{\centering}}
			\hline\hline
			Pole (MeV) &	$(J^P,j)$ &  Channel & \multicolumn{2}{c}{Couplings $g_i \times g_j$}  \\\hline
		2954   & 	$(1/2^-,0)$ &   $\Xi_{c}\bar{K}$   & \multicolumn{2}{c}{$1.912$} \\
		\multirow{2}{*}{3003} &	\multirow{2}{*}{$(1/2^-,1)$} &  \multirow{2}{*}{$\Xi_{c}^\prime\bar{K}$, $\Omega_{c}\eta$}   &   \multirow{2}{*}{$\begin{pmatrix} 3.538& 2.741 \\2.741 & 2.123 \end{pmatrix}$} \\
		& &   \\
		\multirow{2}{*}{3044} &	\multirow{2}{*}{$(3/2^-,1)$} & \multirow{2}{*}{$\Xi_{c}^{*}\bar{K}$, $\Omega_{c}^*\eta$ }   &  \multirow{2}{*}{$\begin{pmatrix}1.731& 1.246 \\1.246 & 0.897 \end{pmatrix}$} \\
		& &  &  \\
		\multirow{2}{*}{3066} &	\multirow{2}{*}{$(1/2^-,1)$} &  \multirow{2}{*}{$\Xi_{c}^\prime\bar{K}$, $\Omega_{c}\eta$}  &  \multirow{2}{*}{$\begin{pmatrix}0.903& 1.257 \\1.257 & 1.750 \end{pmatrix}$} \\
		& &   \\
		\multirow{2}{*}{3113} &	\multirow{2}{*}{$(3/2^-,1)$} & \multirow{2}{*}{$\Xi_{c}^{*}\bar{K}$, $\Omega_{c}^*\eta$ }  &  \multirow{2}{*}{$\begin{pmatrix}2.450& 2.678 \\2.678 & 2.927 \end{pmatrix}$} \\
		& &  \\
		\hline\hline
	\end{tabular*}
\end{center}
\end{table}

It should be mentioned that this assignment is compatible with present measurements including the masses, widths, and spins by the LHCb Collaboration. Also, this picture solves the difficulties of fine structures in the traditional three-quark interpretation and inconsistent numbers in the molecular scenario.  Although the precise results depend on model parameters such as the cutoff parameter $\Lambda$ and bare masses of $\Omega_c(1P_\lambda)$ states, the conclusions of the $\Omega_c(2954)$, $\Omega_c(3000)$ and $\Omega_c(3050)$ states do not change in our coupled-channel perspective. Future BelleII and LHCb experiments can search for the bound states $\Omega_c(2954)$ and measure the spin-parities of particles $\Omega_c(3000)$ and $\Omega_c(3050)$ to test our predictions.

\section{Summary}\label{summary}

In the present work, we have performed a systematic analysis of the low-lying $\Omega_c$ structures in a coupled-channel approach. The couplings between meson-baryon channels, $\Xi_c \bar K$, $\Xi_c^{\prime} \bar K$, $\Xi_c^{*} \bar K$, $\Omega_c\eta$, and $\Omega_c^{*}\eta$, and three-quark bare states $\Omega_c(1P_\lambda)$ have been considered. By solving the Bethe-Salpeter equation, we have found a bound state below the $\Xi_c \bar K$ threshold with a mass of $2954~\rm{MeV}$ and quantum numbers $(J^P,j)=(1/2^-,0)$, which can be searched for in the $\Omega_c^{(*)} \pi$ and $\Omega_c^{(*)} \gamma$ final states. Also, the resonances $\Omega_c(3000)$ and $\Omega_c(3050)$ can be classified as the lower $(J^P,j)=(1/2^-,1)$ and $(J^P,j)=(3/2^-,1)$ states, respectively. Based on present scheme, we have been able to assign the resonances $\Omega_c(3065)$ and $\Omega_c(3090)$ as the conventional $\Omega_{c2}(3/2^-)$ and $\Omega_{c2}(5/2^-)$ states, respectively. The $\Omega_c(3119)$ may correspond to a higher pole with quantum number $(J^P,j)=(3/2^-,1)$. 

Our assignments from this coupled-channel perspective are significantly different from those of traditional three-quark picture and of molecular scenario. Certainly, it should be recognized that present calculations based on phenomenological models inherently carry uncertainties, originating from the approximations and parameters used. Owing to these theoretical uncertainties and unknown spin-parities of the low-lying $\Omega_c$ structures, it is still early to draw the final conclusion for these particles. We hope that future BelleII and LHCb experiments can search for the bound state $\Omega_c(2954)$ and measure the spin-parities of particles $\Omega_c(3000)$ and $\Omega_c(3050)$ to test our predictions.

\subsection*{ACKNOWLEDGMENTS}

We would like to thank Long-Ke Li, Ming-Sheng Liu, Wei Liang, and Ya-Li Shu for helpful discussions. This work is supported by the Natural Science Foundation of  Hunan Province under Grant No. 2023JJ40421, the
Scientific Research Foundation of Hunan Provincial Education Department under Grant No. 24B0063, and the Youth Talent
Support Program of Hunan Normal University under Grant No. 2024QNTJ14. A. Hosaka is supported in part by 
the Grants-in-Aid for Scientific Research [Grants No. 21H04478(A) and No. 24K07050(C)]. H. Nagahiro is supported by the Grants-in-Aid for Scientific Research [Grant No. 24K07051].

\end{document}